\documentclass[twocolumn]{aastex631}
\usepackage{amsmath}
\usepackage{amssymb}
\usepackage{verbatim}
\usepackage{natbib}
\usepackage{txfonts}
\usepackage{graphicx}
\usepackage{url}
\usepackage{multirow}
\usepackage{subfigure}
\usepackage{xcolor}
\usepackage{float}
\restylefloat{table}

\begin{document}

\title{Timing of Seven Isolated Pulsars in the Globular Cluster Terzan~1}

\author[0009-0003-4549-0509]{Justine Singleton}
\affil{Department of Astronomy and Cornell Center for Astrophysics and Planetary Science, Cornell University, Ithaca, NY 14853, USA}
\affil{George Mason University, Department of Physics and Astronomy, Fairfax, VA 22030}
\author[0000-0002-2185-1790]{Megan DeCesar}
\affil{George Mason University, Department of Physics and Astronomy, Fairfax, VA 22030}
%
\author[0000-0002-9618-2499]{Shi Dai}
\affil{Australia Telescope National Facility, CSIRO, Space and Astronomy, PO Box 76, Epping, NSW 1710, Australia}
\author[0000-0002-7965-3076]{Deven Bhakta}
\affil{University of Virginia, Department of Astronomy, P.O. Box 400325, Charlottesville, VA 22904, USA}
\author[0000-0001-5799-9714]{Scott Ransom}
\affil{National Radio Astronomy Observatory, 520 Edgemont Road, Charlottesville, VA 22903, USA}
%
%
\author[0000-0002-1468-9668]{Jay Strader}
\affil{Center for Data Intensive and Time Domain Astronomy, Department of Physics and Astronomy, Michigan State University, East Lansing, MI 48824, USA}
\author[0000-0002-8400-3705]{Laura Chomiuk}
\affil{Center for Data Intensive and Time Domain Astronomy, Department of Physics and Astronomy, Michigan State University, East Lansing, MI 48824, USA}
\author[0000-0003-3124-2814]{James Miller-Jones}
\affil{International Centre for Radio Astronomy Research, Curtin University, GPO Box U1987, Perth, WA 6845, Australia}
%
%

\begin{abstract}

Globular clusters host large populations of millisecond pulsars (MSPs) due to their high gravitational encounter rates, producing many binary systems and thus MSPs via the recycling process. Seven pulsars with spin periods ranging from 3 ms to 134 ms have been discovered in Terzan 1, which was targeted for pulsar searches with the Green Bank Telescope after Australia Telescope Compact Array imaging revealed steep-spectrum point sources in the cluster core. We have obtained timing observations over seven years, for the first seven Green Bank Telescope (GBT) discoveries (Terzan 1 A through G), using the GBT and Murriyang, CSIRO's Parkes radio telescope.
All seven pulsars are isolated, consistent with Terzan 1’s classification as a core-collapsed cluster (core collapse is predicted to disrupt, or ionize, binaries). With these timing solutions,
we measured the positions and observed period derivatives, $\dot{P}$, for each pulsar. The measured $\dot{P}$ values are composed of intrinsic spin-down and accelerations experienced by the pulsars (primarily from the cluster’s gravitational potential), and they can be used to infer line-of-sight accelerations. We attempted to constrain the radius and density of the cluster core using these inferred accelerations. A wide range of radii and densities are possible, pointing to the need for continued timing as well as new discoveries to better constrain these cluster properties.
We additionally find that Ter 1 A may be younger than the cluster and thus may have 
formed via a formation channel other than a core-collapse supernova. Theoretical formation mechanisms such as electron-capture supernovae from accretion- or merger-induced collapse of white dwarfs could potentially explain these pulsars' origins. It may therefore be a member of a small but growing class of globular cluster pulsars that appear to be significantly younger than their host clusters.

\end{abstract}

\section{Introduction \label{sec:intro}}

Millisecond pulsars (MSPs) are recycled pulsars that are spun up by accreted matter from an evolved companion star. While MSP formation in the Galactic field requires the pulsar to already have a companion, more pathways to MSP formation are available in globular clusters (GCs); by mass, MSPs are orders of magnitude more numerous in GCs because the clusters' stellar density increases the number of opportunities for binaries to form \citep[i.e., the encounter rate is high;][]{2014A&A...561A..11V}. Via stellar interactions, especially in the very dense GC core, main sequence (or more evolved) stars can become companions to old neutron stars and subsequently spin them up to millisecond periods \citep{2023MNRAS.520.3847C}.   \par

Timing MSPs in GCs over a long timing baseline can yield insights on cluster dynamics (especially in the cluster core) via precise measurements of the MSPs' spin and orbital periods and derivatives, positions within the cluster, and proper motions.  
Observations of other pulsar parameters also provides measurements of mass distribution throughout clusters \citep{2003MNRAS.340.1359F}.  These allow for improved study of the evolution of GCs.  \par

In this paper, we present timing solutions for the first seven pulsars found in the globular cluster Terzan~1 (hereafter Ter~1).  Ter~1 is a core-collapsed GC in the galactic bulge \citep{2018PASA...35...21P}.  As a core-collapsed GC, it is expected to have a higher single-binary encounter rate and therefore mostly isolated or wide binary pulsars \citep{2014A&A...561A..11V}.  Similarly, it is more likely to have slower, partially recycled pulsars \citep{2022A&A...664A..27R}. 
After steep-spectrum radio sources were detected in Ter~1 with the Australia Telescope Compact Array (ATCA) \citep[MAVERIC Survey;][]{Tudor2022MNRAS.513.3818T}\footnote{The complete catalog from this paper can be found at \url{https://data.csiro.au/collection/csiro:54270}}, these seven pulsars were discovered in the resulting search with the Green Bank Telescope (DeCesar et al. in prep.).
Ter~1~B-G are MSPs, while Ter~1~A has a spin period of $\sim 134$\,ms \footnote{https://www3.mpifr-bonn.mpg.de/staff/pfreire/GCpsr.html}.



This paper is organized as follows. In Section~2, we describe the observations used in this work. In Section~3, we present the results of our timing analyses of the Ter~1 pulsars. We discuss implications of our results, especially uncertainties in the intrinsic spindown rates and cluster-induced accelerations of the pulsars, in Section~4, and draw conclusions in Section~5.

\section{Observations and Data Processing \label{sec:data}}

We obtained $\sim$\,1\,yr of timing data on the Ter~1 pulsars with Murriyang, CSIRO's Parkes radio telescope(hereafter ``Parkes''), and also made use of the discovery and sparsely-spaced follow-up pointings taken with Green Bank Telescope (GBT) between 2017--2023. All timing observations were performed in search mode with coherent dedispersion at the discovery dispersion measure of DM\,$\sim$\,380\,pc\,cm$^{-3}$.
The Parkes observations were all performed with the Ultra-Wideband Low (UWL) receiver (704--4032\,MHz) and the Medusa backend; three initial test observations at Parkes recorded total intensity, while the subsequent pointings recorded full Stokes data. The GBT observations were done at S-band with either the GUPPI or VEGAS backend. A log of the GBT and Parkes timing observations is given in Table~\ref{tab:obsnlog}.



\begin{deluxetable}{ccrc}
\label{tab:obsnlog}
\tablewidth{0pt}
\tabletypesize{\footnotesize}
\small
\tablecaption{Log of Green Bank and Parkes timing observations. All Parkes observations were performed with the Ultra-Wideband Low (UWL) receiver and Medusa backend. $^{a}$The GBT observation on MJD 57855 used the S-band (2\,GHz) receiver with 800\,MHz bandwidth and the GUPPI backend. $^{b}$The GBT observations on MJDs 59645 and 59646 used the S-band receiver with 1500\,MHz bandwidth and the VEGAS backend.}
\tablecolumns{2}
\tablehead{
  \colhead{Telescope} &
  \colhead{Start Date} & \colhead{Start Date} & \colhead{Integration} \\ 
  \colhead{} & \colhead{(YYMMDD)} & \colhead{(MJD)} & \colhead{(s)} \\
}
\startdata
GBT$^{a}$ & 170412 & 57855.3918 & 7855.3 \\
GBT$^{b}$ & 220307 & 59645.4042 & 5571.6 \\
GBT$^{b}$ & 220308 & 59646.4770 & 5584.6 \\
Parkes & 221006 & 59858.2939 & 12392.6 \\ 
Parkes & 221204 & 59917.0463 & 12213.8 \\ 
Parkes & 230206 & 59981.7539 & 12266.8 \\ 
Parkes & 230420 & 60054.8394 & 8690.1 \\ 
Parkes & 230704 & 60129.3786 & 13068.9 \\ 
Parkes & 230805 & 60161.4873 & 12119.4 \\ 
Parkes & 230822 & 60178.4282 & 15091.6 \\ 
Parkes & 230823 & 60179.4008 & 10325.8 \\ 
Parkes & 230828 & 60184.1991 & 6305.6 \\ 
Parkes & 230828 & 60184.4426 & 8592.6 \\ 
Parkes & 230830 & 60186.5515 & 2543.8 \\ 
Parkes & 230902 & 60189.4591 & 8940.7 \\ 
Parkes & 230905 & 60192.4848 & 6543.6 \\ 
Parkes & 230906 & 60193.5365 & 1763.7 \\ 
Parkes & 230909 & 60196.4635 & 7266.6 
\enddata
\end{deluxetable}


We processed the GBT and Parkes data using standard {\sc PRESTO}\footnote{https://github.com/scottransom/presto} routines to zap radio frequency interference (RFI), subband and dedisperse the data, and fold the data to detect each pulsar. With the initial Parkes observations, we performed standard pulsar searches (acceleration searches with low or zero $z_{\mathrm{max}}$, as the pulsars are isolated). 

\section{Results \label{sec:results}}



\subsection{Timing Analysis}

Starting from re-detections of pulsars A through G in acceleration searches, we folded all the Parkes observations with the same spin period for a given pulsar and allowed PRESTO's {\tt prepfold} routine to search in P-Pdot space. We then refined these folds to prepare for TOA generation.


\begin{deluxetable*}{lcc}
\tablewidth{0pt}
\tabletypesize{\footnotesize}
\small
\tablecaption{Timing parameters of Ter~1 pulsars.}
\tablecolumns{3}
\tablehead{
  \colhead{Parameter} & \colhead{Ter\,1\,A} & \colhead{Ter\,1\,B}
}
\startdata
\multicolumn{3}{c}{Measured Parameters} \\
\hline
Right ascension (J2000) & 17:35:47.225(3) & 17:35:47.1959(2)\\ 
Declination (J2000) & -30:28:55.2(3) & -30:28:54.94(2)\\
Spin frequency (Hz) & 7.47972799962(4) & 89.79376997879(4)\\
Spin frequency derivative (Hz\,s$^{-1}$) & $-4.8635(3) \times 10^{-15}$ & $7.21908(4) \times 10^{-14}$\\ 
Dispersion measure (pc\,cm$^{-3})$ & 383.20 & 380.67\\ 
Reference epoch (MJD) & 60184.250 & 57855.000\\
Span of timing data (MJD) & 57855-60196& 57855-60196\\
Number of TOAs & 205& 208\\
rms timing residual ($\mu$ s) & 485.60 & 44.26\\
\hline
\multicolumn{3}{c}{Derived Parameters} \\ 
\hline
Spin period $P$ (ms) & 133.6947 & 11.13663\\
Spin period derivative $\dot{P}$ (s\,s$^{-1}$) & $8.69 \times 10^{-17}$ & $-8.95 \times 10^{-18}$\\ 
\enddata
\label{tab:params}
\end{deluxetable*}

\begin{deluxetable*}{lcc}
\tablewidth{0pt}
\tabletypesize{\footnotesize}
\small
\tablecaption{Timing parameters, continued.}
\tablecolumns{3}
\tablehead{
  \colhead{Parameter} & \colhead{Ter\,1C} & \colhead{Ter\,1D}
}
\startdata
\multicolumn{3}{c}{Measured Parameters} \\
\hline
Right Ascension (J2000) & 17:35:47.4558(4) & 17:35:47.1872(5) \\ 
Declination (J2000) & -30:28:46.42(4) & -30:28:57.61(5) \\
Spin frequency (Hz) & 165.5951965811(1) & 185.6537144892(1)\\
Spin frequency derivative (Hz\,s$^{-1}$) &$1.215(1) \times 10^{-15}$ & $-9.5817(1) \times 10^{-14}$\\
Dispersion measure (pc cm$^{-3}$) & 379.850 & 380.720 \\
Reference epoch (MJD) & 59981.750 & 60184.250\\
Span of timing data (MJD) & 57855-60196& 57855-60196\\
Number of TOAs & 120 & 207\\
rms timing residual ($\mu$ s) & 61.61 & 101.69\\
\hline
\multicolumn{3}{c}{Derived Parameters} \\ 
\hline
Spin period $P$ (ms) & 6.038822 & 5.386372\\
Spin period derivative $\dot{P}$ (s\,s$^{-1}$) & $-4.43 \times 10^{-20}$ & $2.77 \times 10^{-18}$\\
\enddata
\label{tab:params2}
\end{deluxetable*}

\begin{deluxetable*}{lccc}
\tablewidth{0pt}
\tabletypesize{\footnotesize}
\small
\tablecaption{Timing parameters, continued.}
\tablecolumns{4}
\tablehead{
  \colhead{Parameter} & \colhead{Ter\,1E} & \colhead{Ter\,1F} & \colhead{Ter\,1G}
}
\startdata
\multicolumn{4}{c}{Measured Parameters} \\
\hline
Right Ascension (J2000) & 17:35:47.2889(2) & 17:35:47.1763(6) & 17:35:48.1547(6)\\ 
Declination (J2000) & -30:28:56.47(2) & -30:28:56.94(6) & -30:28:57.62(5)\\
Spin frequency (Hz) & 324.8476352266(1) & 191.7997850567(2) &  255.2397431635(2)\\
Spin frequency derivative (Hz\,s$^{-1}$) & $-5.9060(1) \times 10^{-14}$ & $7.5753(2)\times 10^{-14}$ & $-9.026(2) \times 10^{-15}$\\
Dispersion measure (pc cm$^{-3}$) & 378.730 & 381.840 & 382.000\\
Reference epoch (MJD) & 59981.751323 & 60184.250 &  60184.452\\
Span of timing data (MJD) & 57855-60196& 57855-60196& 57855-60196\\
Number of TOAs & 88 & 54 & 67\\
rms timing residual ($\mu$ s) & 45.11 & 55.42 & 108.17\\
\hline
\multicolumn{4}{c}{Derived Parameters} \\ 
\hline
Spin period $P$ (ms) & 3.0783 & 5.21377 & 3.917885\\
Spin period derivative $\dot{P}$ (s\,s$^{-1}$) & $5.59 \times 10^{-19}$ & $-2.05 \times 10^{-18}$ & $1.38 \times 10^{-19}$\\
\enddata
\label{tab:params3}
\end{deluxetable*}

TOAs were generated using the {\tt get\_TOAs.py} command in {\tt PRESTO}.  We used {\tt TEMPO}\footnote{https://sourceforge.net/projects/tempo/} to find phase-connected timing solutions for the pulsars with a solar system ephemeris of DE440.  For Ter~1~A-G, we were able to phase-connect the Parkes TOAs backward in time to the Green Bank TOAs from the pulsars' discovery and early follow-up observations, resulting in a longer timing baseline and a longer lever arm for measuring timing parameters, especially astrometric parameters.

\subsection{Timing Results}

The results for these pulsars are listed in Tables 2--4. We measure the spin period derivative $\dot{P}$ for all.
We tested the timing solutions for proper motion but did not detect it. 
\par

We attempted to match the pulsars' timing positions with the positions of ATCA radio sources, but the timing positions are not yet precise or unique enough to perform a one-to-on match with the ATCA-detected sources.


We have not included derived characteristic ages $\tau_c$, magnetic fields, or spin-down energies in the timing parameter tables because they depend on the pulsars' intrinsic spin-down rates $\dot{P}_{\mathrm{int}}$. These rates are unknown due to the pulsars' accelerations, which often cause the observed $\dot{P}$ to differ significantly from the intrinsic value. Therefore the true $\tau_c$ (and other derived parameters) may be very different from the apparent $\tau_c$. The effects of accelerations on $\dot{P}$ are discussed further in the next section.

\section{Discussion}

\subsection{Contributions to observed $\dot{P}$ values}


The observed $\dot{P}$ values, $\dot{P}_{\mathrm{obs}}$, are composed of an unknown combination of intrinsic spindown ($\dot{P}_{\mathrm{int}}$) and several sources of accelerations experienced by the pulsar:

\begin{equation}
\left ( \frac{\dot{P}}{P} \right) _{\mathrm{obs}} = \left ( \frac{\dot{P}}{P} \right) _{\mathrm{int}} + \frac{a_{\mathrm{GC}} + a_{\mathrm G} + a_{\mathrm{PM}} + a_{\mathrm{NN}}}{c},
\end{equation}

\noindent where
$(\dot{P}/P)_{\mathrm{obs}}$ and $(\dot{P}/P)_{\mathrm{int}}$ are the ratios between the spin period and spindown rate for the observed and intrinsic values respectively,
$a_{\mathrm{GC}}$ is the pulsar's acceleration due to the GC potential,
$a_{\mathrm G}$ is the pulsar's acceleration due to the Galactic potential,
$a_{\mathrm{PM}}$ is the centrifugal acceleration due to the transverse Doppler effect from the pulsar's proper motion, known as the Shklovskii effect \citep{Shklovskii1970SvA....13..562S}, $a_{\mathrm{NN}}$ is acceleration from nearest neighbors, and $c$ is the speed of light.

The fractional contribution of each is very difficult to disentangle, especially without binary parameters, but we can obtain analytical order-of-magnitude estimates of these values to determine which acceleration terms are most important. (Typically, the cluster acceleration term dominates over the others.)

\subsubsection{Acceleration in the cluster potential \label{subsubsec:cluster_accel}}

As in \citet{Phinney1993ASPC...50..141P} and \citet{Lynch2012ApJ...745..109L}, the pulsar's maximum acceleration in a core-collapsed GC can be estimated from the central velocity dispersion. We therefore replace $a_{\mathrm{GC}} \rightarrow \max{a_{\mathrm{GC}}}$ and use the expression:

\begin{equation}
\frac{\max{a_{\mathrm{GC}}}}{c} = \frac{3\sigma_v^2}{2c(R_c^2 + R_p^2)^{1/2}}
\end{equation}

\noindent where $\sigma_v$ is the central velocity dispersion, $R_c$ is the radius of the cluster core, and $R_p$ is the projected distance of the pulsar from the cluster center. We took the values of these parameters from the Baumgardt catalog\footnote{https://people.smp.uq.edu.au/HolgerBaumgardt/globular/parameter.html} \citep{Baumgardt2018MNRAS.478.1520B}. For all of the pulsars, we find $\max{a_{\mathrm{GC}}/c} \approx 3$--$4 \times 10^{-17}$\,s$^{-1}$.


\subsubsection{Acceleration in the Galactic potential \label{subsubsec:galactic_accel}}

The cluster's, and thus the cluster pulsars', acceleration due to the Galactic potential can be estimated as \citep{Nice1995ApJ...441..429N}:

\begin{equation}
    \frac{a_{\mathrm G}}{c} = -\frac{\cos{b}}{c} \left ( \frac{\Theta_0^2}{R_0} \right ) \left ( \cos{l} + \frac{\beta}{\sin^2{l} + \beta^2} \right )
\end{equation}

Here $b$ and $l$ are the Galactic latitude and longitude of the pulsar (or of the cluster center), $R_0$ is the Sun's distance from the Galactic Center, $\Theta_0$ is the Galactic rotational velocity at $R_0$, $\beta \equiv D/R_0$,  and $D \approx 5670$\,pc \citep{Baumgardt2018MNRAS.478.1520B} is the distance from the Sun to Ter~1. Using $\Theta_0$ and $R_0$ from \citet{Reid2014ApJ...783..130R}, we find $a_G/c \approx 1.5 \times 10^{-18}$, an order of magnitude lower than the cluster-induced acceleration.



\subsubsection{Shklovskii effect \label{subsubsec:shklovskii}}

$a_{\mathrm{PM}}$ is the centrifugal acceleration due to the transverse Doppler effect from the pulsar's proper motion, known as the Shklovskii effect \citep{Shklovskii1970SvA....13..562S}:

\begin{equation}
\frac{a_{\mathrm{PM}}}{c} = \frac{\mu^2 D}{c}
\end{equation}

\noindent where $\mu$ is the magnitude of the proper motion (in units of angular distance per time, i.e. radians per second). We find $a_{\mathrm{PM}}/c \approx 4 \times 10^{-19}$ for all the Ter~1 pulsars. Thus the Shklovskii effect is also negligible compared to the acceleration in the cluster potential.

\subsubsection{Acceleration due to nearby objects}

We use Equation~4 of \citet{Dai2023MNRAS.521.2616D} to calculate the nearest-neighbor acceleration $a_{\mathrm{NN}}$ that the Ter~1~A pulsars may experience:

\begin{align}
\frac{a_{\mathrm{NN}}}{GM_{\mathrm{cl}}/R_{\mathrm{cl}}^2} \sim 10^{-2} \left ( \frac{10^6}{N} \right ) ^{1/3},
\end{align}


\noindent where $M_{\mathrm{cl}}$ and $R_{\mathrm{cl}}$ are the mass and radius of some portion of the cluster, and $N$ is the number of stars in that part of the cluster. Since the pulsars are in or near the core, we choose to calculate $a_{\mathrm{NN}}$ with the core parameters, such that $M_{\mathrm{cl}} \rightarrow M_{\mathrm c} \sim \rho_{\mathrm c} r_{\mathrm c}^3$ and $R_{\mathrm{cl}} \rightarrow r_{\mathrm c}$. We use the values of these parameters from \citet{Baumgardt2018MNRAS.478.1520B}, $r_{\mathrm c} \approx 0.36$\,pc and $\rho_{\mathrm c} \approx 63000 M_{\odot}$\,pc$^{-3}$. For $N_{\mathrm c} = M_{\mathrm c}/<M_*>$, where $<M_*>$ is the average mass of stars in the cluster, we assume a mean mass of $0.45 M_{\odot}$ based on the stellar types remaining in Ter~1 \citep[K and M types;][]{Valenti2015A&A...574A..80V}. We find $a_{\mathrm{NN}}/c = 1.5\times 10^{-18}$\,s$^{-1}$. This is smaller than $a_{\mathrm{GC}}$ by an order of magnitude and thus can be ignored.



\subsection{Modeling cluster properties with acceleration profiles \label{subsec:accelprof}}

We created an acceleration profile to compare the pulsars' accelerations to their observations.  The measured $\dot{P}$ values result from an unknown combination of the above accelerations, especially acceleration in the cluster potential, and spin-down.  For binary pulsars, it would be possible to use measurements of changes in orbital period to constrain the contributions of acceleration or spin-down \citep[e.g.,][]{Prager2017ApJ...845..148P}, but as all the known pulsars in Terzan 1 are isolated, it is not possible to disentangle the effects due to each acceleration from effects due to spin-down.  We begin by assuming that the observed $\dot{P}$ values are from acceleration and the effect of spin-down is negligible.  A line-of-sight pseudo-acceleration, 
\begin{equation}
a_{\mathrm{los}} = c \frac{(\dot{P}_{\mathrm{obs}} - \dot{P}_\mathrm{int})}{P},
\label{eqn:a_los}
\end{equation}
\noindent can then be used to approximate the pulsars' true line-of-sight accelerations. This psuedo-acceleration $a_{\mathrm{los}}$ was plotted against the projected angular separation between the cluster center and each pulsar (again, under the assumption that the intrinsic spin-down rate $\dot{P}_\mathrm{int}$ is negligible). The angular separation values were calculated based on the pulsar positions, which were determined from timing, and the cluster center, as catalogued by Baumgardt$^2$.
\par

To model the acceleration profile of Ter~1 with $a$ of Ter~1~A--G, we used Equation (27) from \citet{Prager2017ApJ...845..148P}, based on Equation (3.5) from \citet{Phinney1993ASPC...50..141P}:  
\begin{equation}
a_1 = \frac{2 \pi G \rho_c r_c^2}{\sqrt{r_c^2 + R_\perp^2}} 
\end{equation}
where $R_\perp$ is the projected angular separation between the cluster center and a pulsar's timing position.  Using the core radius, $r_c$, and the core density, $\rho_c$, as variables, we plotted a model for each combination of the two which closely encompassed all seven pulsar psuedo-acceleration values.  
\par

\begin{figure}
    \centering
    \includegraphics[width=\linewidth]{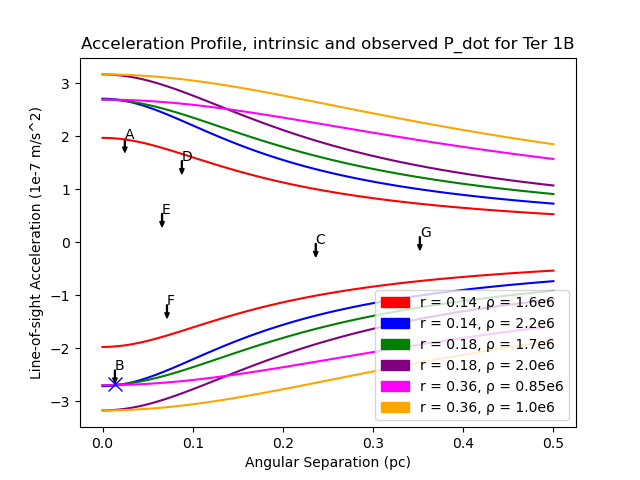}
    \caption{Acceleration profiles for Ter 1 as calculated in \citet{Prager2017ApJ...845..148P}, with the upper limits of the line-of-sight accelerations $a_1 = c\dot{P}/P$ of Ter~1~A--G (black arrows) plotted against the angular separation. These accelerations are upper limits because we have not accounted for the (unknown) intrinsic spin-down rates of the pulsars. We additionally plot a large-magnitude but realistic value of $a_1$ for Ter~1~B (blue ``X''), assuming an intrinsic spin-down rate of $\dot{P}_{\mathrm{int}} = 1 \times 10^{-18}$ because pulsar B has the most constraining $a_1$. Each curve represents a possible fit for our model. The three curves (blue, green, and magenta) passing through the realistic $a_1$ of pulsar B demonstrate that there is a wide range of $r_{\mathrm c}$ and $\rho_{\mathrm c}$ that produce the plotted $a_1$. Longer-term timing of these pulsars and of future additional pulsar discoveries will further constrain the core radius and density, e.g. via measurements of proper motion or of orbital period derivatives if one or more binary pulsars are discovered in Ter~1.}
    \label{fig:accelprofile}
\end{figure}

Figure~\ref{fig:accelprofile} shows several acceleration profiles to demonstrate how the radius and density parameters impact the fit.  Because we are treating the pulsars' intrinsic spin-down rates as negligible, the accelerations plotted here represent upper limits on the true accelerations.
As can be seen in Figure~\ref{fig:accelprofile}, $a$ from Ter~1~B provides the strongest constraints on $r_c$ and $\rho_c$. We therefore explore how the cluster parameters change when we include a large but feasible value of intrinsic spindown ($\dot{P}_{\mathrm{B}} = 1\times 10^{-18}$),
based on properties of pulsars within the known pulsar population with similar $P$, for Ter~1~B. 
We find that a wide range of core radii and mass densities are consistent with the $a_{\mathrm{los}}$ of pulsar B; more pulsars will be needed to constrain the cluster radius and density.

The Baumgardt catalog estimates a core radius of $r_c = 0.36$ pc and the logarithm of the core density as $\log(\rho_c) = 4.80$ $M_\sun/$pc$^{3}$, or $\rho_c \approx 63095$ $M_\sun/$pc$^{3}$. For our curve with $r_\mathrm{c} = 0.36$\,pc, we find a density $\sim 8.5e5\,M_{\odot}$\,pc$^3$, roughly consistent with that of the Baumgardt catalog. 
It is possible that more pulsars will be discovered in Terzan 1, and timing those could further constrain the core radius.

\subsection{Empirical constraints on cluster acceleration and estimates of intrinsic spindown}

Pulsars B, C, and F have $\dot{P}_{\mathrm{obs}}<0$ and thus appear to be spinning up. Their spin derivatives must therefore be significantly contaminated by their acceleration, the majority of which is due to their line-of-sight acceleration in the cluster potential  \citep[this is typical of cluster MSPs, e.g.,][]{Prager2017ApJ...845..148P}, with the negative value of acceleration indicating motion toward the Earth.
The other pulsars with $\dot{P}_{\mathrm{obs}}>0$ are also experiencing cluster acceleration, either to a lesser degree (toward Earth) in the Earth-pulsar line of sight such that their $\dot{P}_{\mathrm{obs}}$ remains positive, or in the opposite direction (away from Earth) such that their $\dot{P}_{\mathrm{obs}}$ remains positive and is larger than the intrinsic spin-down rate.

With the values of $P$ and $\dot{P}_{\mathrm{obs}}$ for pulsars B, C, and F, we can estimate $a_{\mathrm{los}}$ for these three pulsars using Equation~\ref{eqn:a_los}.
%
%
We note that $a_{\mathrm{los}}$ is dominated by acceleration in the cluster potential, as demonstrated in Section~\ref{subsubsec:cluster_accel}. We do not know the intrinsic spin-down rate, $\dot{P}_{\mathrm{int}}$, so we can only estimate a range of values for $a_{\mathrm{los}}$ by assuming values of $\dot{P}_{\mathrm{int}}$. 
In the most conservative case, we assume no spin-down, i.e. $\dot{P}_{\mathrm{int}} = 0$, which will yield a lower limit on $a_{\mathrm{los}}$.To find a realistic upper limit on the acceleration, we use a relatively high yet plausible value of $\dot{P}$ (based on measured properties of the general MSP population), $\dot{P}_{\mathrm{int}} = 10^{-18}$. Table~\ref{tab:inferred_accel} lists the inferred boundary values of $a_{\mathrm{los}}/c$.

\begin{deluxetable}{ccccc}
\tablewidth{0pt}
\tabletypesize{\footnotesize}
\small
\tablecaption{Inferred line-of-sight accelerations of MSPs with $\dot{P}_{\mathrm{obs}} < 0$, assuming $0 < \dot{P}_{\mathrm{int}} < 10^{-18}$.}
\label{tab:inferred_accel}
\tablecolumns{5}
\tablehead{
  \colhead{Pulsar} & \colhead{$P$ (s)} & \colhead{$\dot{P}_{\mathrm{obs}}$} & \colhead{Assumed $\dot{P}_{\mathrm{int}}$} & \colhead{Inferred $a_{\mathrm{los}}/c$ (s$^{-1}$)}
}
\startdata
B & 0.01113663 & -8.9534e-18 & 0.0 & $-8.040 \times 10^{-16}$ \\
 & & & $10^{-18}$ & $-8.938 \times 10^{-16}$ \\[1.5ex]
C & 0.00603882 & -4.4314e-20 & 0.0 & $-7.338 \times 10^{-18}$ \\
 & & & $10^{-18}$ & $-1.729 \times 10^{-16}$ \\[1.5ex] 
F & 0.00521377 & -2.0592e-18 & 0.0 & $-3.950 \times 10^{-16}$ \\
 & & & $10^{-18}$ & $-5.868 \times 10^{-16}$ 
\enddata
\end{deluxetable}

We can now explore what fraction of the observed $\dot{P}$ for the other pulsars could be due to acceleration rather than to the pulsars's intrinsic spindown.
%
We use the largest-magnitude acceleration value in Table~5, $a_{\mathrm{los}}/c = -8.9 \times 10^{-16}$\,s$^{-1}$ (from Ter~1B), to place upper and lower bounds on $\dot{P}_{\mathrm{int}}$ for pulsars A, D, E, and G. The resulting maximum and minimum values of $\dot{P}_{\mathrm{int}}$
are in Table~6. In each case, we find $\min{(\dot{P}_{\mathrm{int}})} < 0$; therefore we can only place meaningful upper limits on the intrinsic spin-down rates of these pulsars.

\begin{deluxetable}{ccccc}
\tablewidth{0pt}
\tabletypesize{\footnotesize}
\small
\tablecaption{Estimated upper and lower bounds on $\dot{P}_{\mathrm{int}}$, assuming $a_{\mathrm{los}}/c = -8.9 \times 10^{-16}$\,s$^{-1}$ from Ter~1~B in Table~5.}
\tablecolumns{5}
\tablehead{
  \colhead{Pulsar} & \colhead{$P$ (s)} & \colhead{$\dot{P}_{\mathrm{obs}}$} & \colhead{$\max{(\dot{P}_{\mathrm{int}})}$} & \colhead{$\min{(\dot{P}_{\mathrm{int}})}$}
}
\startdata
A & 0.1336947 & $8.6933 \times 10^{-17}$ & $2.07 \times 10^{-16}$ & $<0$ \\
D & 0.005386372 & $2.77 \times 10^{-18}$ & $7.59 \times 10^{-18}$ & $<0$\\
E & 0.0030783 & $5.59 \times 10^{-19}$ & $3.31 \times 10^{-18}$ & $<0$ \\
G & 0.003917885 & $1.38 \times 10^{-19}$ & $3.64 \times 10^{-18}$ & $<0$
\enddata
\end{deluxetable}

\subsection{Possible age and formation of Ter~1~A and Ter~1 MSPs}

The age of Ter~1 is roughly 12\,Gyr: a recent age estimate is 12.6$\pm$1.3\,Gyr \citep{Kharchenko2015yCat..35850101K, Kharchenko2016A&A...585A.101K}, while combining results from \citet{Gontcharov2021MNRAS.508.2688G} and \citet{Ortolani1999A&A...350..840O} yields an age of 11.5$\pm$1.1\,Gyr. Comparing the cluster's age with the range of possible ages inferred for Ter~1~A, we find that Ter~1~A may be significantly younger than the cluster, but we cannot draw this conclusion definitively at this point. It could have a very high intrinsic spin-down rate and have formed very recently; if its true intrinsic spin-down rate were $\max{(\dot{P}_\mathrm{int})} = 2\times10^{-16}$ from Table~6, then its characteristic age would be only $\sim$\,11\,Gyr. However, Ter~1~A could instead have a very low intrinsic spin-down rate and have been formed by a core-collapse supernova early in the cluster's history.



\citet{2014A&A...561A..11V} tie the encounter rate to the likelihood of exchange or ionization interactions occurring between a binary pulsar and another star, producing new binaries or isolated pulsars. They suggest that seemingly young globular cluster pulsars may be old, partially-recycled pulsars that lost their companions in such interactions. However, \citet{Kremer2024arXiv240907527K} argue that this partial recycling scenario is unlikely, because the time needed to recycle a pulsar to slow periods like that of Ter~1~A is too short for a gravitational encounter that ionizes the binary to statistically occur.

If Ter~1~A is in fact significantly younger than Ter~1, then it is possible that instead of forming from a type II core-collapse supernova of a massive star, it formed through an as-of-yet theoretical formation mechanism, particularly the accretion- or merger-induced collapse of a white dwarf leading to an electron-capture supernova.  In the case of an accretion-induced collapse, matter from a companion star is accreted onto a degenerate white dwarf until it reaches critical mass and collapses into a pulsar.  Merger-induced collapse also involves a binary, but instead of one white dwarf, there are two which coalesce and merge \citep{2008MNRAS.386..553I}.  Others have previously conducted research on this being a mechanism for neutron star formation in globular clusters, particularly in globular clusters with young pulsars \citep{2008MNRAS.386..553I, 2019ApJ...877..122Y}.  There has also been speculation about specific young pulsars in globular clusters forming from electron-capture supernovae, such as PSR NGC6440A \citep{2008ApJ...675..670F} and PSR J1823-3021C in NGC 6624 \citep{Lynch2012ApJ...745..109L}.

The fact that all of the pulsars found thus far in Ter~1 are isolated is consistent with predictions of binary ionization in core-collapsed clusters. More recent work has found that rather than binary disruption, the large fraction of MSPs in globular clusters may instead be due to tidal disruption of main sequence stars by neutron stars \citep[and subsequent accretion that spins up the pulsar;][]{Kremer2022ApJ...934L...1K, Ye2022ApJ...934L...1K}, or the merger and subsequent collapse of binary white dwarfs. NGC~6752 is another example of a core-collapsed cluster with a large population of isolated MSPs where these mechanisms may operate; and localizations of fast radio bursts in extragalactic globular clusters \citep{Bhardwaj2021ApJ...910L..18B, Kirsten2022Natur.602..585K} or similar galactic environments \citep{Eftekhari2024arXiv241023336E} also support the possibility of alternative formation mechanisms for globular cluster MSPs in addition to slow pulsars.


\section{Conclusions}

In this paper, we presented results of long-term timing of the first seven pulsars discovered in Terzan 1.
All the pulsars are isolated, which is consistent with expectations for a core-collapsed globular cluster.  


The observed spin-down values for the pulsars are the result of an unknown combination of intrinsic spin-down and accelerations in the cluster potential, in the Galactic potential, and from the Shklovskii effect.  Because the pulsars are all isolated, the effects of spin-down and accelerations cannot be disentangled from each other.  Using the upper limits of the pulsars' line-of-sight accelerations $a_1$ together with a large-magnitude but feasible $a_1$ for Ter~1~B (calculated by assuming a relatively large intrinsic spin-down rate of $1\times 10^{-18}$), we plotted a range of acceleration profiles that are parameterized with the core radius and density of the cluster. We find that we cannot constrain $r_{\mathrm c}$ and $\rho_{\mathrm c}$ with our current data.


Ter 1B, Ter 1C, and Ter 1F have negative spin-down rates, a sign of significant contamination from acceleration.  To estimate the magnitude of this acceleration, we use Equation 7 and assume values of $\dot{P}_{\mathrm{int}}$ to constrain upper and lower limits for the accelerations.  We then use the maximum estimate for the acceleration to constrain $\dot{P}_{\mathrm{int}}$ for the remaining pulsars.


We find that Ter 1A could be as young as $\sim$\,11 Myr, 
while the age of Ter 1 is estimated to be $\sim$\,12 Gyr. Alternatively, it may be as old as the cluster and have been partially recycled before losing its companion in a gravitational encounter.
If young, then Ter~1~A would be part of a small but growing class of globular cluster pulsars that are significantly younger than their host cluster, supporting the existence of neutron formation mechanisms aside from only core-collapse supernovae of massive, short-lived stars.  Theoretical formation mechanisms such as electron-capture supernovae from accretion-induced collapse or merger-induced collapse of a white dwarf could potentially explain these pulsars' origins.  


\section*{Acknowledgments}

The Green Bank Observatory
is a facility of the NSF operated under cooperative agreement by Associated Universities, Inc. The National Radio Astronomy Observatory is a facility of the NSF operated under cooperative agreement by Associated Universities, Inc. 
The Parkes radio telescope (Murriyang) is part of the
Australia Telescope National Facility which is
funded by the Australian Government for operation as a National Facility
managed by CSIRO. We acknowledge the Wiradjuri People as the traditional
owners of the Observatory site. We acknowledge the Wallumedegal People
of the Darug Nation and the Wurundjeri People of the Kulin Nation as the
traditional owners of the land where this work was carried out.
M.D. and J. Singleton acknowledge support from the National Science Foundation (NSF) Physics Frontiers Center award No. 2020265.
J. Singleton additionally acknowledges support from Cornell University's work-study program.
J. Strader acknowledges support from NASA grant 80NSSC21K0628 and NSF grant AST-2205550.
Radio astronomy research in L.C.'s group at Michigan State is supported by NSF grants AST-2107070 and AST-2205628.





\bibliographystyle{aasjournal}
\bibliography{ter1_timing_js_noMeerKAT}

\end{document}